\documentclass[preprint]{acmart}
\AtBeginDocument{%
  }

\setcopyright{acmlicensed}
\copyrightyear{2018}
\acmYear{2018}
\acmDOI{XXXXXXX.XXXXXXX}
\acmConference[Conference acronym 'XX]{Make sure to enter the correct
  conference title from your rights confirmation email}{June 03--05,
  2018}{Woodstock, NY}
\acmISBN{978-1-4503-XXXX-X/2018/06}
\usepackage{tabu}                      
\usepackage{booktabs}                  
\usepackage{lipsum}                    
\usepackage{mwe}                       
\usepackage{tabularx}
\usepackage{float}
\usepackage{stfloats}
\usepackage{placeins}

\usepackage{amsmath}
\usepackage{mathptmx}                  
\usepackage{subcaption}

\begin{document}
\newcommand{\ww}{WaveWalker}
\newcommand{\wwc}{WaveWalkerClone}
\title{Exploration of Radar-based Obstacle Visualizations to Support Safety and Presence in Camera-Free Outdoor VR}

\author{Avinash Ajit Nargund}
\email{anargund@ucsb.edu}
\orcid{0009-0005-4224-3240}
\affiliation{%
  \department{Electrical and Computer Engineering}
  \institution{University of California, Santa Barbara}
  \city{Santa Barbara}
  \state{CA}
  \country{USA}
}

\author{Andrew L. Huard}
\email{ahuard@ucsb.edu}
\affiliation{%
  \department{Electrical and Computer Engineering}
  \institution{University of California, Santa Barbara}%
  \city{Santa Barbara}
  \state{CA}
  \country{USA}}

\author{Tobias Höllerer}
\email{holl@cs.ucsb.edu}
\orcid{0000-0002-6240-0291}
\affiliation{%
  \department{Computer Science}
  \institution{University of California, Santa Barbara}
  \city{Santa Barbara}
  \state{CA}
  \country{USA}
}

\author{Misha Sra}
\email{sra@cs.ucsb.edu}
\orcid{0000-0001-8154-8518}
\affiliation{%
  \department{Computer Science}
  \institution{University of California, Santa Barbara}
  \city{Santa Barbara}
  \state{CA}
  \country{USA}
}

\renewcommand{\shortauthors}{Nargund et al.}

\begin{abstract}
Outdoor virtual reality (VR) places users in dynamic physical environments where they must remain aware of real-world obstacles, including static structures and moving bystanders, while immersed in a virtual scene. This dual demand introduces challenges for both user safety and presence. Millimeter-wave (mmWave) radar offers a privacy-preserving alternative to camera-based sensing by detecting obstacles without capturing identifiable visual imagery, yet effective methods for communicating its sparse spatial information to users remain underexplored. In this work, we developed and validated WaveWalkerClone, a reproduction of the WaveWalker system \cite{nargund2024wavewalker}, to establish reliable radar- and GPS–IMU–based sensing under varied outdoor lighting conditions (n = 9). Building on this feasibility validation, we conducted a user study (n = 18) comparing three visualization techniques for radar-detected obstacles: (1) diegetic alien avatars that visually embed obstacles within the virtual narrative, (2) non-diegetic human avatars represented obstacles as humans inconsistent with the virtual narrative, and (3) abstract point clouds centered around the obstacles conveying spatial data without anthropomorphic or narrative associations. Our results show that all three approaches supported user safety and situational awareness, but yielded distinct trade-offs in perceived effort, frustration, and user preference. Qualitative feedback further revealed divergent user responses across conditions, highlighting the limitations of a one-size-fits-all approach. We conclude with design considerations for obstacle visualization in outdoor VR systems that seek to balance immersion, safety, and bystander privacy.
\end{abstract}


\keywords{Do, Not, Use, This, Code, Put, the, Correct, Terms, for,
  Your, Paper}
\begin{teaserfigure}
  \centering
      \includegraphics[width=\textwidth]{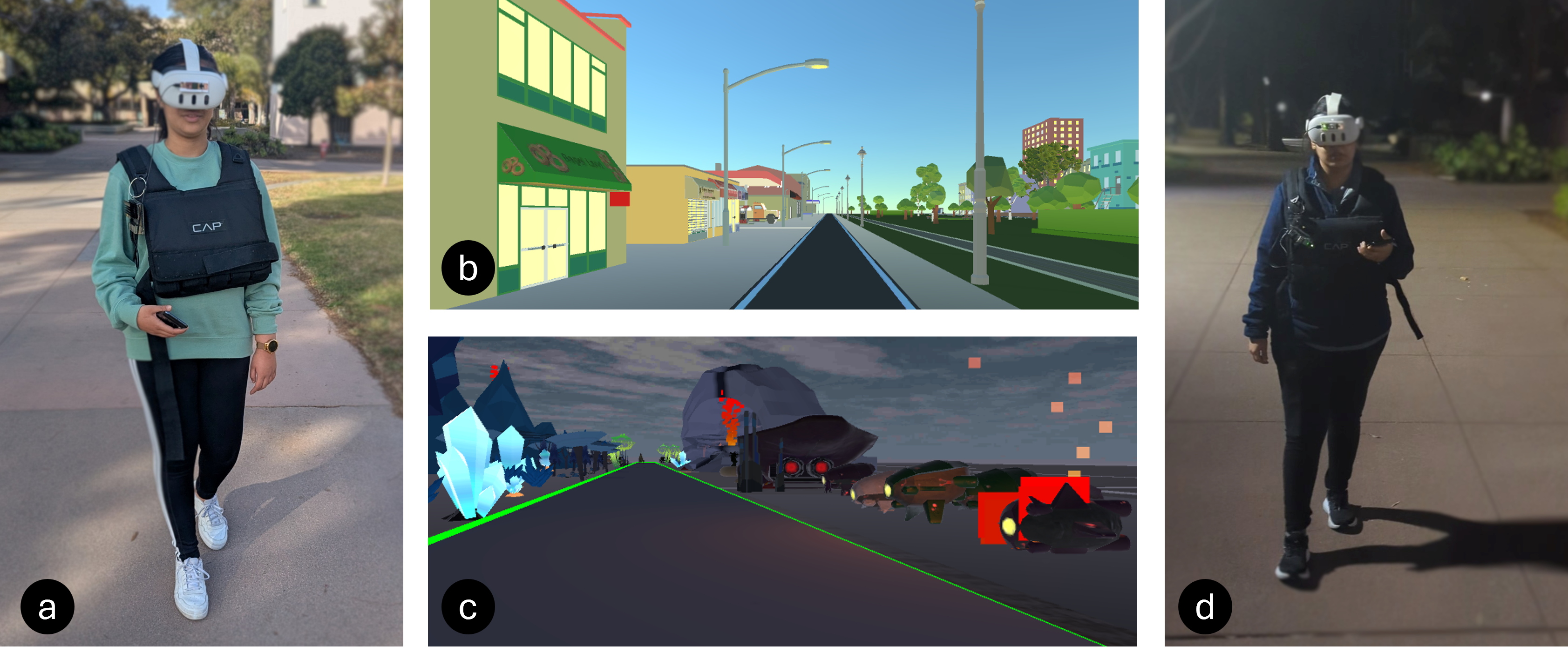}
 \caption{We investigate how mmWave-detected bystanders and obstacles can be visualized to support safety, situational awareness, and presence in outdoor VR. We first validated a camera-free sensing pipeline using \ww~\cite{nargund2024wavewalker} under varied lighting conditions (a, d) in a city-themed environment (b). Building on this validation, we conducted a visualization study in an extraterrestrial-themed environment (c), comparing three obstacle representations: diegetic alien avatars, non-diegetic human avatars, and abstract point clouds. Results reveal distinct trade-offs in user experience, underscoring the need for adaptable visualization strategies in mmWave-based outdoor VR systems.}
  \label{fig:teaser}
\end{teaserfigure}

\received{20 February 2007}
\received[revised]{12 March 2009}
\received[accepted]{5 June 2009}

\maketitle

\section{Introduction}

Outdoor virtual reality (VR) experiences, such as walking-based applications (e.g., DreamWalker~\cite{yang2019dreamwalker}), promise new forms of interactive engagement. While indoor VR supports activities such as gaming or guided tourism, it is constrained by room-scale boundaries and often relies on artificial locomotion techniques, which can diminish a user's sense of presence \cite{slater1995taking}. Outdoor VR leverages full-scale environments to enable more natural movement, offering a potential solution to the challenge of VR locomotion \cite{al2018virtual} and advancing the vision of hybrid-reality experiences \cite{Ritterbusch2023}.

Delivering safe outdoor VR requires addressing challenges in tracking, spatial awareness, and bystander privacy. Commercial VR headsets typically rely on camera-based tracking and pass-through video, but these methods are unreliable in varied outdoor conditions (e.g., changing daylight or fog), have limited tracking range (e.g., 10–15 feet on the Meta Quest 3), and raise privacy concerns for bystanders \cite{do2023vice, hadan2024privacy}. To overcome these limitations, outdoor VR systems may benefit from alternative sensors. For this work, we created \textit{WaveWalkerClone}, a replication of the WaveWalker system~\cite{nargund2024wavewalker}, which combines GPS–IMU fusion for localization with millimeter-wave (mmWave) radar for detecting static and dynamic obstacles. While mmWave addresses many drawbacks of cameras (e.g., lighting, weather, range, or privacy), it produces sparse spatial data, raising the open question of how best to visualize obstacle information for users.

Prior work has explored various techniques for representing real-world obstacles in VR, including overlays such as avatars, maps, or composited models \cite{von_willich_you_2019, yang2019dreamwalker, Kudo2021, gottsacker_diegetic_2021}, as well as visual encodings such as point clouds and depth grids \cite{kanamori_obstacle_2018, huang_improving_2018}. These strategies illustrate different ways of balancing safety and immersion, but research has largely focused on indoor or camera-based systems. Extending them to outdoor VR requires rethinking visualization under sparse, camera-free sensing constraints.


In this work, we first validated \textit{WaveWalkerClone} under different outdoor lighting conditions to confirm that the mmWave–GPS–IMU pipeline could reliably detect nearby bystanders without compromising user safety or immersion. Building on this system feasibility step, our primary study compared three visualization techniques for depicting mmWave detected bystanders within a fictional alien VR environment. The techniques differ in whether the visual representation is shown as an entity that belongs to the world's fiction or as a representation that is perceptually present but thematically incongruent. To characterize this distinction, we draw on the concept of diegesis from game and media studies. Following Galloway~\cite{galloway2006gaming}, we define diegetic elements as those that are consistent with the internal logic and narrative framing of the virtual environment, and non-diegetic elements as those that violate that framing. Using this definition, we tested a diegetic alien avatar, a non-diegetic human avatar, and an abstract point cloud baseline.

Our work makes the following contributions:
\begin{itemize}
    \item An empirical validation of \textit{WaveWalkerClone}, demonstrating that mmWave radar combined with GPS–IMU tracking functions reliably under varied outdoor lighting conditions and has the potential to support safe outdoor VR. 
    \item A comparative evaluation of three radar-based obstacle visualization techniques, revealing trade-offs between safety, presence, and immersion in outdoor VR.
    \item Design considerations for integrating obstacle information into privacy-preserving outdoor VR systems, grounded in user preferences and experiences with the mmWave-based \textit{WaveWalkerClone}
\end{itemize}

In sum, our work takes a step toward making outdoor VR both safe and immersive while preserving bystander privacy. By validating WaveWalkerClone and comparing obstacle visualization techniques, we provide initial evidence and design considerations for radar-based systems that support privacy-preserving outdoor VR, laying groundwork for future hybrid-reality experiences.

\section{Related Work}
Our work builds on prior research in enhancing awareness of the tracked area boundary as well as static and dynamic obstacles in the user's immediate surroundings to support safe and immersive VR experiences. 

\subsection{User Awareness of Surroundings}

A core challenge in VR is that users are isolated from their physical environment, creating risks of collisions with boundaries, furniture, or bystanders \cite{OHagan2020a}. Prior work has addressed this through two major approaches: supporting awareness of play-area boundaries and representing obstacles within the tracked space.

Consumer systems such as the HTC Vive's Chaperone and Meta Quest's Guardian provide cues when users approach the boundary of a predefined play area, either by displaying a grid or switching to pass-through video. Research prototypes extend this idea through redirected walking \cite{razzaque2005redirected, fan_redirected_2023}, resets or distractors \cite{peck2010improved, sra2018vmotion}, and alternative sensory channels like audio or haptics \cite{sra2016procedurally, george2020invisible} to steer the user away from the boundary. While effective indoors, these techniques assume fixed boundaries and may not generalize well to outdoor environments, where space is often unbounded and obstacles are unpredictable.

Beyond fixed boundaries, users must also be aware of both static obstacles, like furniture, and dynamic obstacles, particularly bystanders who may enter the space unexpectedly, introducing additional challenges due to their unpredictability and movement \cite{OHagan2020a, Kudo2021}. To address this, bystander awareness was enabled through selective integration of real-world elements into VR to facilitate safe interaction without overwhelming the user \cite{mcgill2015dose}. However, alternate approaches show that even abstract visualizations can effectively alert users to nearby obstacles or bystanders while maintaining immersion \cite{ghosh2018notifivr, sra2017oasis}. 

While prior work has explored safety in VR through boundary awareness, obstacle detection, bystander notification systems, and inclusion of real-world geometry into VR, these efforts have largely been limited to simulated public settings or static, pre-mapped indoor spaces with stable lighting. Less is known about how obstacle representations should be designed in outdoor VR contexts, where sensing is sparse, obstacles may be dynamic, and bystander privacy precludes the use of cameras as a sensing device. Building on insights from these indoor studies, we investigate bystander and obstacle awareness in outdoor VR through visual representations of static and dynamic obstacles detected using \textit{WaveWalkerClone} to help enhance a user's sense of safety.

\subsection{Obstacle Visualization Strategies}
To provide obstacle and bystander awareness, researchers have developed several distinct visualization strategies. Techniques include substituting real objects with virtual proxies \cite{simeone2015substitutional,valentini2020improving}, overlaying pre-scanned geometry \cite{huang_improving_2018, kanamori_obstacle_2018}, compositing live RGB-D captures into the virtual world \cite{hartmann2019realitycheck}, or blending depth-based representations of nearby objects \cite{Chung2023}. These techniques communicate real-world obstacles to the user but can often disrupt immersion or raise privacy concerns when applied outdoors due their reliance on dense visual sensing.

Other work has explored more abstract encodings, such as point clouds or grids, to provide a minimal but effective spatial awareness of obstructions \cite{liu2021virtual}. While such representations support safety, they may not integrate naturally into the narrative world of the virtual environment. Conversely, representational avatars or proxies for real-world objects can enhance coherence and engagement \cite{simeone2015substitutional, valentini2020improving}, but it remains unclear how they influence perceived safety when obstacles are dynamic or unpredictable.

Overall, these works show the range of visualization techniques available for conveying environmental hazards. However, they have been studied primarily in indoor, camera-based contexts. How abstract versus representational approaches should be designed for outdoor VR, where sensing is sparse and privacy-preserving, remains largely unexplored.
\begin{figure}[!t]
\centering
\includegraphics[scale=0.45]{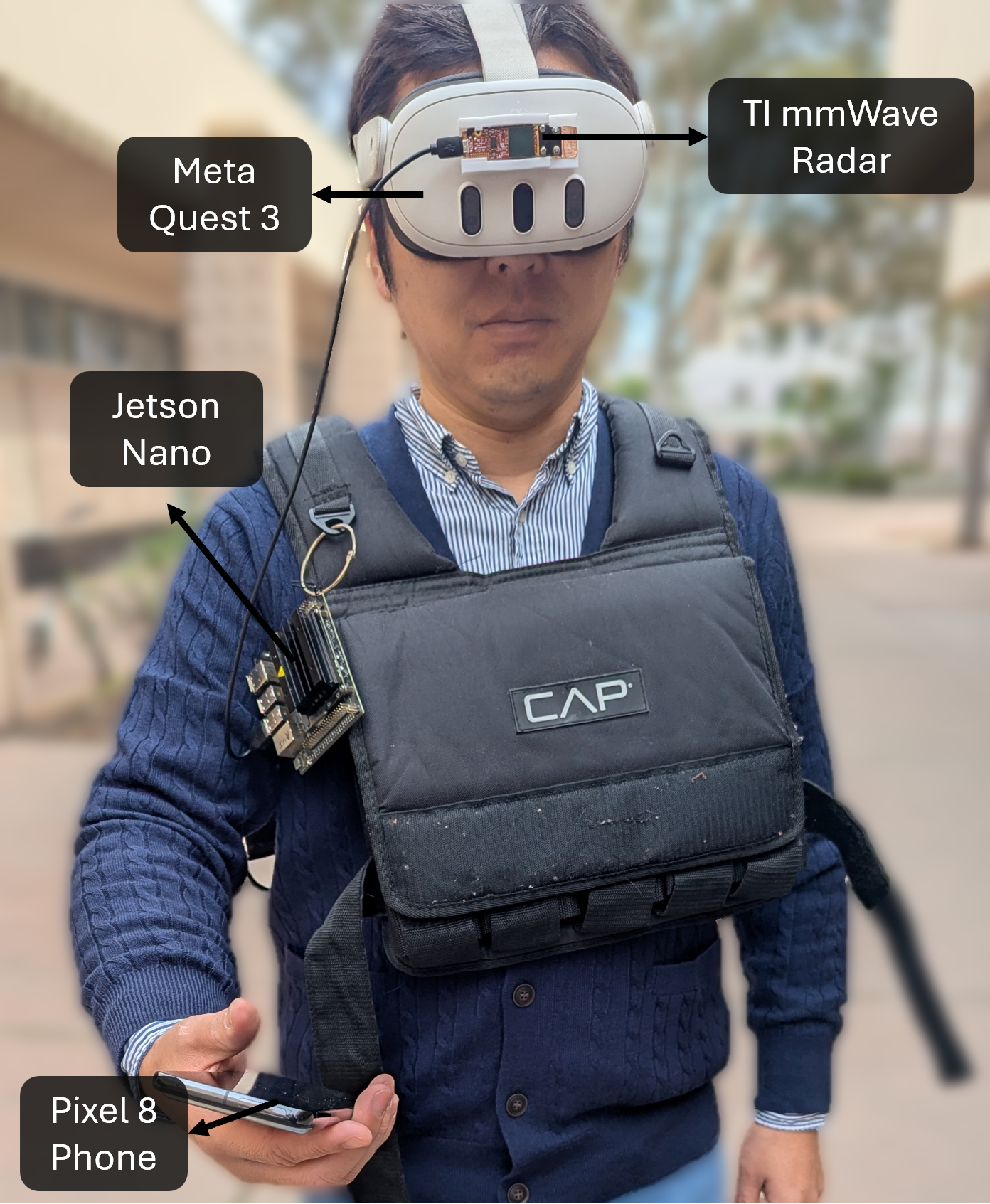}
\caption{A user wearing the \textit{WaveWalkerClone} outdoor VR system which comprises of a Pixel 8 smartphone which provides GPS and IMU data. A Jetson Nano fuses the GPU-IMU data, processes TI mmWave radar data, and streams output to the Meta Quest 3 headset.}
\label{fig:ww_sys}
\end{figure}
\subsection{Diegetic Representations}

The concept of diegesis relates to elements that belong to the narrative world and has been adopted in VR to describe interface cues that appear as natural parts of the virtual environment rather than as overlays or external signals \cite{galloway2006gaming, salomoni2016assessing}. Diegetic representations have been shown to strengthen presence and coherence in applications such as cinematic VR and redirected walking \cite{rothe2019guidance, cao2020automatic, sra2018vmotion}. Reinforcing this, Gottsacker et al. \cite{gottsacker_diegetic_2021} tested five different interruption notifications affected user experience, with the key finding that users preferred diegetic notifications the most.

Prior work suggests that diegetic cues can enhance immersion, but their role in safety-critical contexts is less clear. In particular, little is known about how diegetic versus non-diegetic approaches affect user perception of safety and spatial awareness when representing static or dynamic obstacles from real-world environments in VR. Most existing studies focus on entertainment or guidance tasks, leaving open how visualization strategies should be designed when the goal is to keep users safe during locomotion. Our study is motivated by the specific trade-off between the narrative benefits of diegesis and the functional requirements of spatial awareness in outdoor VR.

While prior work has established techniques for boundary awareness, obstacle visualization, and diegetic design in VR, these approaches have primarily been studied in indoor, camera-based settings. At the same time, though diegetic design is known to enhance immersion, its specific role in enhancing a user's sense of safety remains under explored in VR research. Our work, to the best of our knowledge, is one of the first to empirically compare diegetic, non-diegetic, and abstract obstacle visualizations in such a context, using mmWave radar as a privacy-preserving sensing modality.

\section{System Design}

\begin{figure*}[!t]
\centering
\includegraphics[scale=0.6]{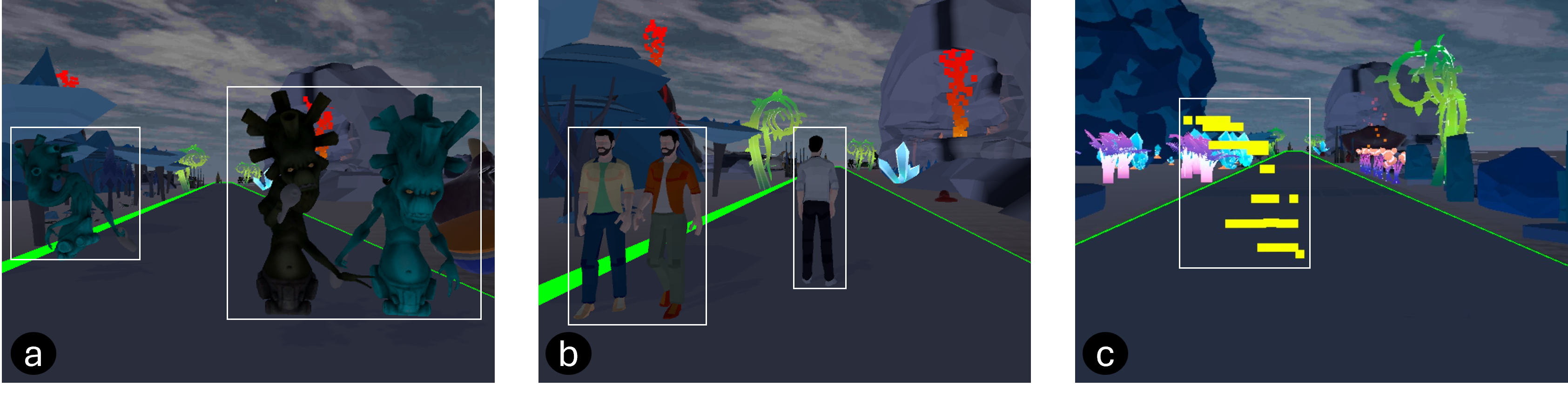}

\caption{In our visualization study, participants explored an extraterrestrial-themed VR environment featuring real-world bystanders rendered using three visualization strategies: (a) diegetic alien avatars, (b) non-diegetic human avatars, and (c) abstract point clouds. Bystander representations as seen in VR by the study participant are outlined here with bounding boxes for clarity.}
\label{fig:viz_types}
\end{figure*}

To evaluate visualization techniques for outdoor VR, we required a camera-free system capable of detecting obstacles in real-world environments. We created \textit{WaveWalkerClone}, a replication of the original WaveWalker system~\cite{nargund2024wavewalker}, which combines GPS–IMU fusion for localization with millimeter-wave (mmWave) radar for detecting static and dynamic obstacles.
This setup allows outdoor operation without continuous video capture, preserving bystander privacy.  

Figure~\ref{fig:ww_sys} illustrates the components used in our implementation:  
\begin{itemize}
\setlength\itemsep{0em}
  \item \textbf{Google Pixel 8:} Provides GPS and IMU data.  
  \item \textbf{Texas Instruments IWR6843AOP mmWave radar:} Detects static and dynamic objects.  
  \item \textbf{Meta Quest 3 HMD:} Equipped with the mounted radar module.  
  \item \textbf{NVIDIA Jetson Nano:} Fuses GPS–IMU data via an Error-State Kalman Filter (ESKF)~\cite{sola2017quaternion}, processes radar data, and streams output to the headset.  
\end{itemize}

The radar module connects to the Jetson via USB-Serial, while other components communicate over Wi-Fi using ROS \cite{quigley2009ros}. A user's pose updates are published at 20~Hz and obstacle detections at 10~Hz. We selected these two rates based on outdoor pilot testing. This system served as the enabling platform for our studies. In Study~1, we evaluated \textit{WaveWalkerClone's} functionality under variable lighting conditions to establish feasibility for safe outdoor use. In Study~2, we built on this validation to investigate how different visualization techniques could convey obstacle information to enhance a user's sense of safety.

\subsection{VR Application}

We developed our VR application in Unity\footnote{\texttt{https://unity.com/}} (version 2022.3) using the OpenXR\footnote{\texttt{https://www.khronos.org/openxr/}} backend. Using the point cloud visualization from \ww~\cite{nargund2024wavewalker} as a baseline, we introduced two additional visualization modes of diegetic and non-diegetic avatars to examine how different strategies affect user safety and immersion.

For the system validation (Study~I), we constructed a VR environment whose spatial layout aligned 1:1 with a section of our campus, using 3D building and path data from OpenStreetMap\footnote{\texttt{https://www.openstreetmap.org/}} as the base. We overlaid this layout with city-themed Unity assets such as buildings, pedestrians, vehicles, and trees to create an immersive urban environment (Figure~\ref{fig:teaser}(b)). Obstacles detected by the mmWave radar were rendered as point clouds within this environment, allowing us to evaluate the feasibility of \textit{WaveWalkerClone} for outdoor VR under varying outdoor lighting conditions.


For the visualization study (Study~II), we designed a VR environment set on an alien planet while preserving the underlying campus spatial layout to maintain alignment between the virtual and physical spaces (Figure~\ref{fig:teaser}(c)). Participants were tasked with physically walking through the environment to reach a spaceship located approximately 200m away, while safely navigating around nearby bystanders detected by our radar sensing system. Walkable paths and physical landmarks were retained to support locomotion, whereas buildings and non-traversable regions were replaced with thematically coherent elements such as futuristic structures, large boulders, and dense alien vegetation. We introduced a narrative in which the user is traversing an unfamiliar alien world to reach their spacecraft, providing an in-world rationale for the environment and the presence of non-human entities. This narrative framing provided a consistent in-world context for interpreting obstacle representations during movement and avoidance. Within this setting, we examined how different visualization techniques for nearby bystanders influenced perceived safety and sense of presence.
Specifically, we evaluated three visualization conditions (Figure~\ref{fig:viz_types}): 

1. \textit{Alien avatars} (diegetic): Bystanders were represented as alien characters whose appearance, motion, and behavior were consistent with the alien-world setting.
2. \textit{Human avatars} (non-diegetic): Bystanders were represented as human figures in contemporary casual clothing, visually incongruent with the alien environment.
3. \textit{Point clouds} (abstract baseline from \ww~\cite{nargund2024wavewalker}): Bystanders were represented as dynamic point clouds indicating position and movement without narrative or character cues.

\begin{figure*}[!t]
\centering
\includegraphics[scale=0.5]{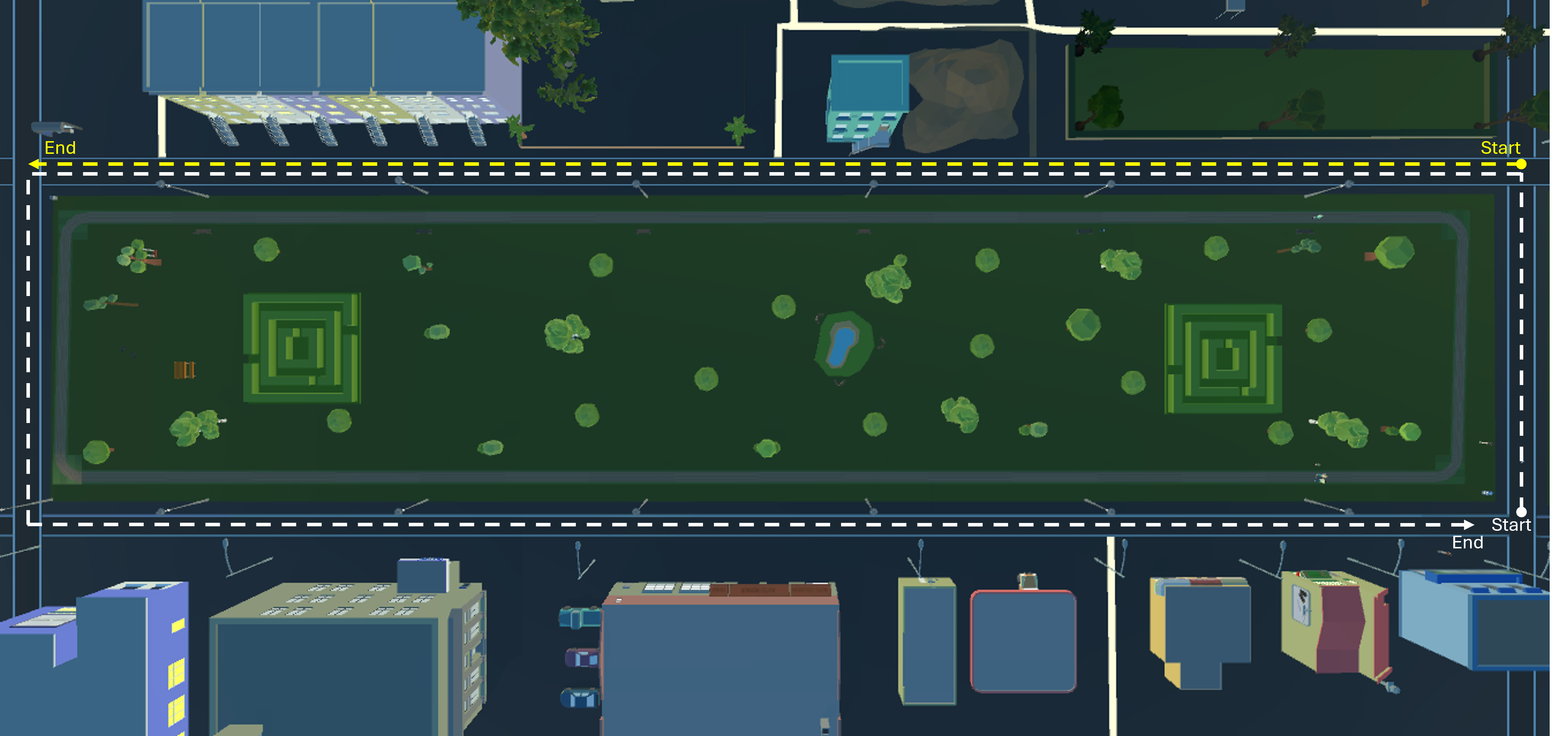}

\caption{Aerial view of the campus quad used as the basis for the outdoor VR environment in the city theme used in Study~I. In the preliminary system evaluation, participants walked approximately 500 m around the full quad (white dashed line). In the visualization study, participants followed a shorter 200m route along one side (yellow dashed line), beginning at the circle and ending at the arrow tip.}
\label{fig:ve_route}
\end{figure*}

\subsection{Visualization Implementation}
To translate radar detections into meaningful cues in VR, we implemented a set of rules for how obstacles were tracked, instantiated, and maintained in the virtual environment. The design goals were twofold: (i) ensure visualizations are responsive and can support outdoor safety and (ii) provide clear signals of both static and dynamic obstacles. 

A key design decision was to represent obstacles as avatars rather than abstract markers. Prior work has shown that approaching avatars in VR elicits measurable physiological responses in users, such as increased skin conductance, which heightens awareness of nearby movement \cite{Llobera2010}. We therefore expected avatars to provide a more salient signal of obstacle presence compared to abstract encodings, particularly for dynamic obstacles. Point clouds were retained as an abstract baseline to benchmark against.

We defined an $8 m \times 6 m$ Region of Interest (ROI) around the user, corresponding to the radar's effective range (slightly over 8 m) and the paved path width (±3 m per side). Only obstacles within this ROI were tracked and visualized. Constraining the ROI in this way helped reduce spurious detections beyond useful walking distance while still covering the area most relevant to avoid collisions.

Through extensive outdoor pilots, we tuned radar tracking parameters to minimize false negatives and achieved a track creation latency, which is the duration from an obstacle entering the ROI to the creation of its track, of 0.4–0.6 seconds. This latency is acceptable for walking-speed encounters ($1.2$ to $1.4$ m/s \cite{bohannon1997comfortable}) as even with a conservative closing speed of 2.8 m/s (both user and bystander moving),  a user covers about 1.5 m in this time window, providing adequate distance for avoidance. Moreover, human visual reaction time averages around $250$ ms \cite{shelton2010comparison}, so the total distance a user moves in the combined system-plus-human response time remains well within the 8 m ROI, ensuring safety.

All obstacles within the ROI were treated uniformly, including pedestrians, cyclists, and static objects (such as cones or bicycles), due to \ww's \cite{nargund2024wavewalker} use of a DBSCAN clustering-based object detector, which we replicated in \textit{WaveWalkerClone}. While this detector cannot provide semantic labels or orientations, detailed classification was not essential for our task. Prior work on proxy objects \cite{simeone2015substitutional} suggests that signaling whether a space is occupied is often sufficient for collision avoidance.

Visualization behavior was tailored to condition. In the point cloud condition, each tracked object was rendered using radar points accumulated over three frames (0.3s), which reduced jitter without introducing perceptible lag. In the avatar conditions, static obstacles (including stationary people) were shown as idle avatars with subtle animations (e.g., head turns, breathing), to mitigate the ``frozen statue'' effect that prior VR work has shown to disrupt plausibility \cite{rothe2019guidance}. For dynamic obstacles, we prioritized trajectory accuracy while adapting locomotion behavior to maintain narrative consistency. Non-diegetic human avatars walked with full-body walking animations to mirror real-world gait, whereas the diegetic alien avatar glided smoothly along the path. 
This contrast allowed us to test how narrative consistency versus representational clarity influenced user experience \cite{galloway2006gaming}.

Lastly, we implemented persistence rules to reduce visual clutter and confusion. When an object exited the ROI, its avatar remained in place as though it had stopped moving while still displaying subtle idle animations. Expired tracks behind the user were gradually faded out. Smooth fade-outs were chosen over instant removal to prevent abrupt ``popping'' effects, consistent with blending techniques shown to preserve immersion in prior spatial safety visualizations \cite{hartmann2019realitycheck}.

\begin{figure*}[h]
\centering
\begin{subfigure}{0.24\textwidth}
\centering
\includegraphics[width=\linewidth]{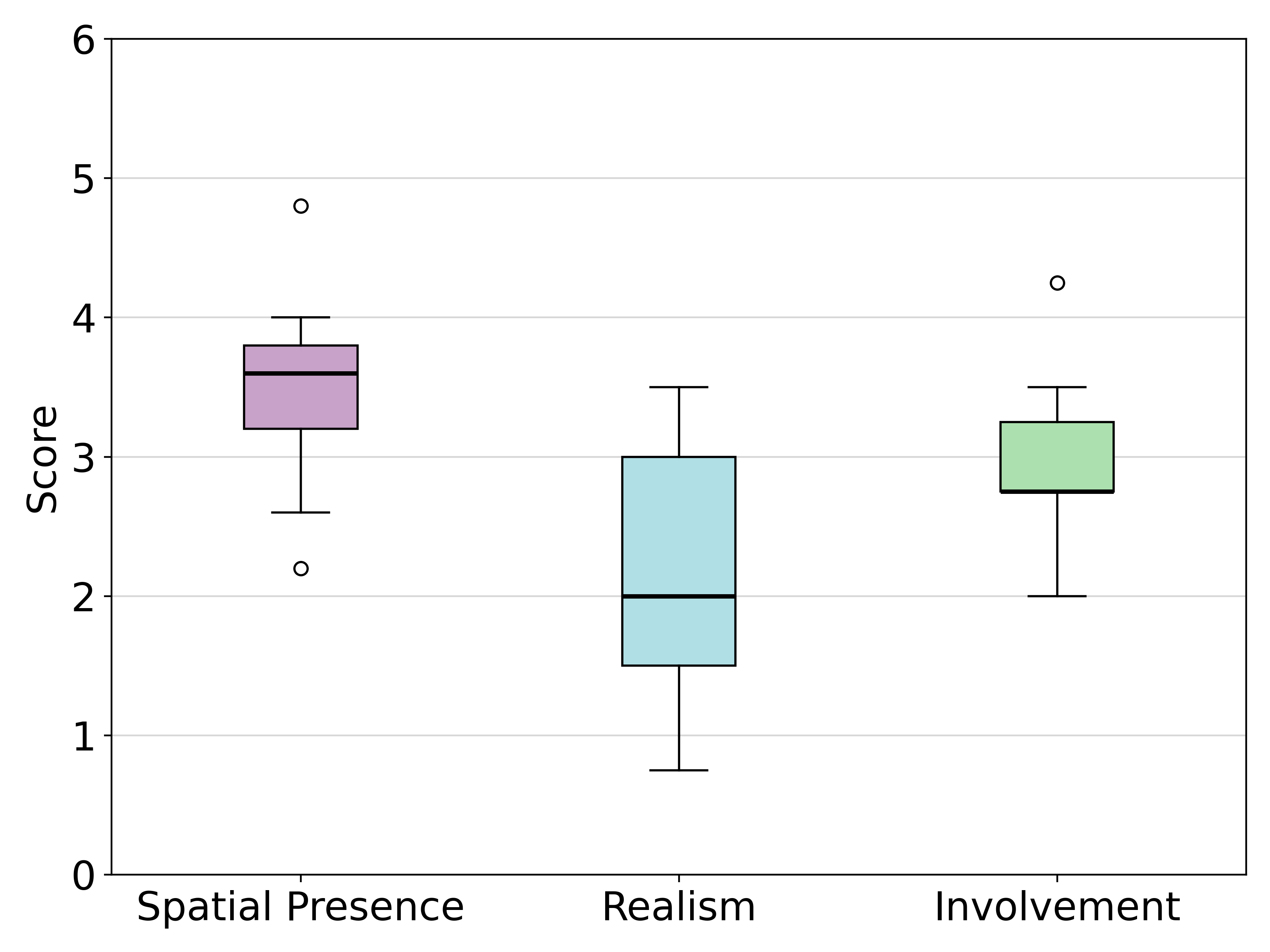}
\caption{IPQ Scores}
\label{fig:ipq}
\end{subfigure}
\begin{subfigure}{0.24\textwidth}
\centering
\includegraphics[width=\linewidth]{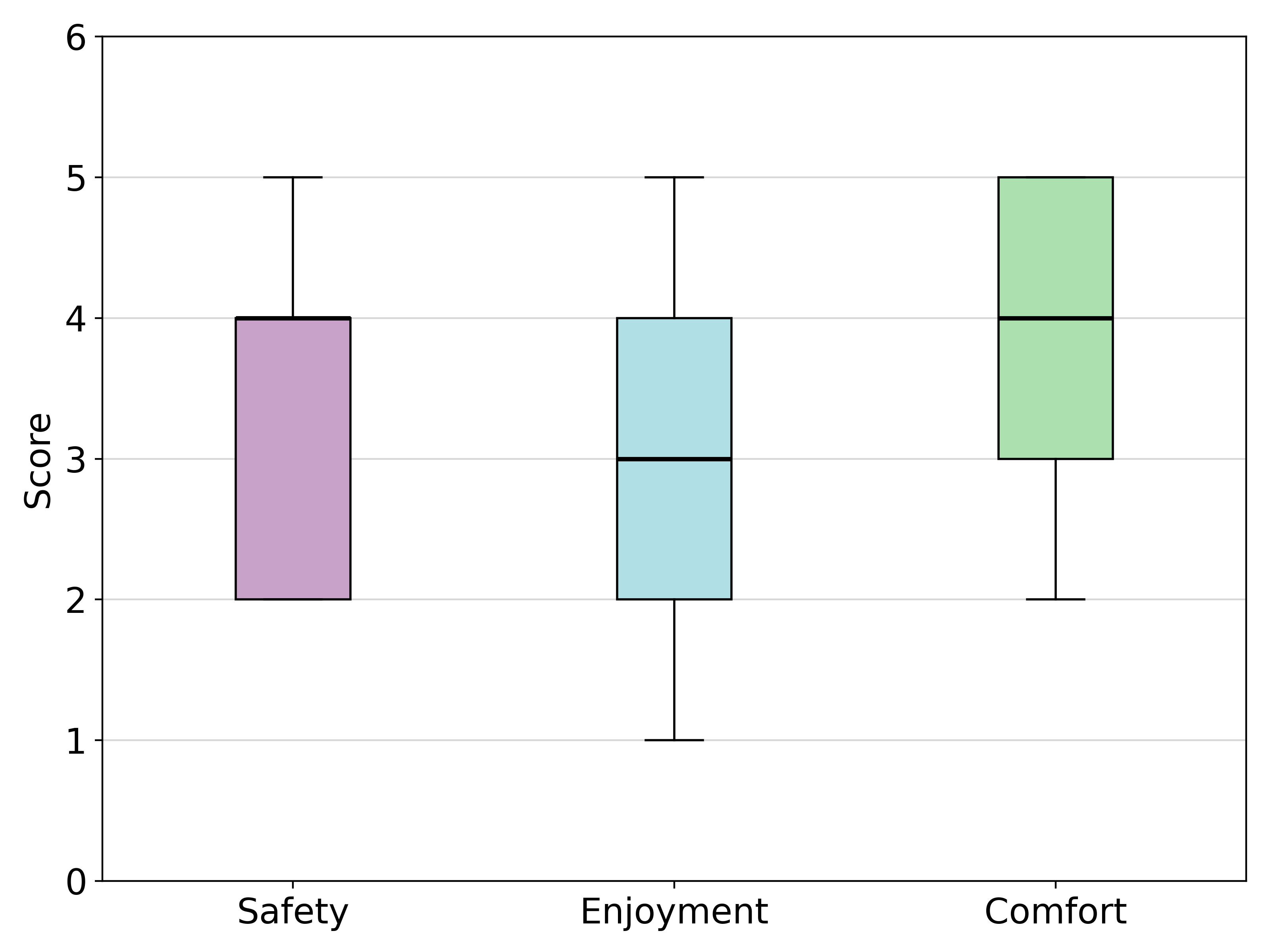}
\caption{Enjoyment and comfort}
\label{fig:psq}
\end{subfigure}
\begin{subfigure}{0.24\textwidth}
\centering
\includegraphics[width=\linewidth]{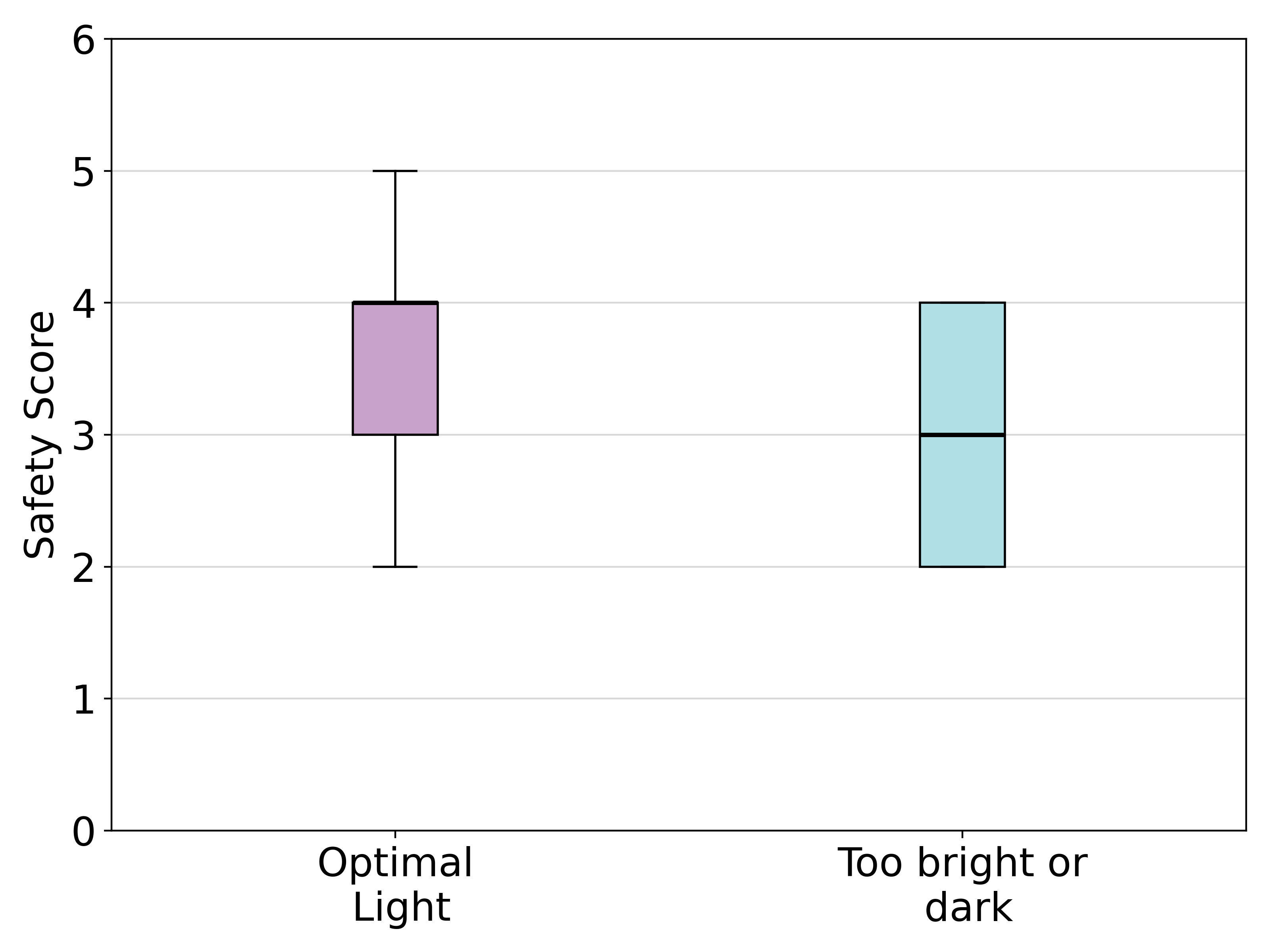}
\caption{Perceived safety score}
\label{fig:sl}
\end{subfigure}
\begin{subfigure}{0.24\textwidth}
\centering
\includegraphics[width=\linewidth]{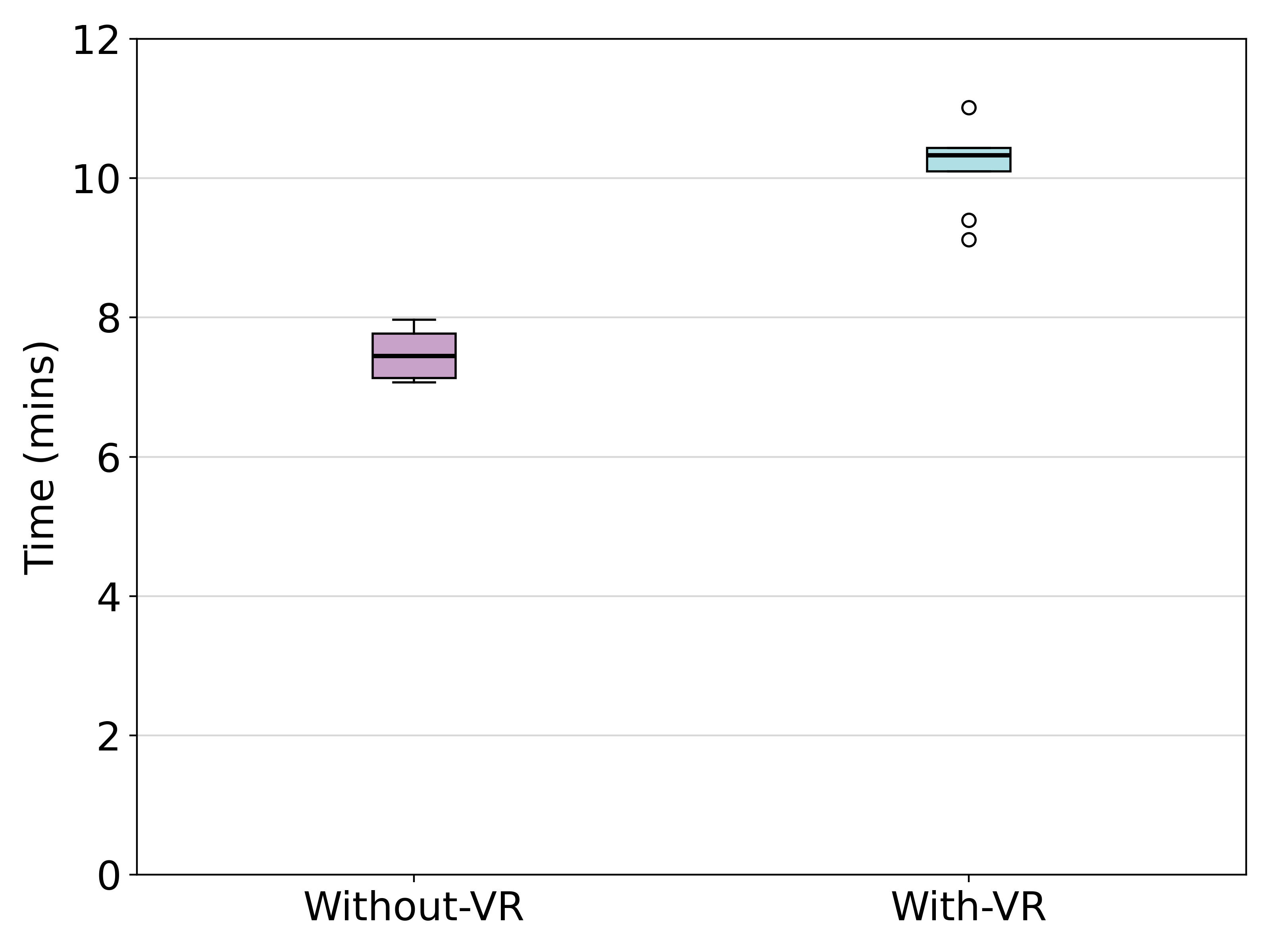}
\caption{Walking times}
\label{fig:wt}
\end{subfigure}
\caption{This figure presents the results from our preliminary system evaluation (Study~I). The plots show (a) scores for spatial presence, realism and involvement scales of the IPQ, (b) ratings for perceived safety, enjoyment, and comfort, (c) perceived safety across different lighting conditions, and (d) a comparison of walking times between the with-VR and without-VR conditions.}
\label{fig:eval_res}
\end{figure*}

\section{User Study I : System Evaluation}
\label{section:se}
We first conducted a small-scale feasibility study to evaluate the safety and comfort of our \textit{WaveWalkerClone} system in an outdoor setting (N=9). The study site was a busy campus quad with consistent foot traffic but no automobile traffic, chosen to approximate the characteristics of public pedestrian spaces such as parks, plazas, and sidewalks. We aimed to assess whether the system functioned reliably under realistic outdoor conditions, including varied lighting and unpredictable bystander movement.

\subsection{Participants}
We recruited nine participants (5 male, 3 female, 1 non-binary; ages 18–24, median age 19). Four had limited VR experience, while three were frequent users (gaming or coursework). All were compensated \$20 for their time, and the study was approved by our local IRB (protocol \# anonymized). To assess the sensor's functionality outdoors, participants completed the study under varied outdoor lighting conditions: bright sun (n=2), dusk/darkness (n=2), drizzle (n=1), and mild daylight (n=4).

\subsection{Procedure}
Participants provided informed consent and completed a demographic questionnaire before the study. They were briefed on the walking route (Figure~\ref{fig:ve_route}) and shown the point cloud obstacle visualization before starting the with-VR condition. This was the only visualization used in Study~I. Each participant completed two counterbalanced laps (one with VR, one without) around the campus quad. After the with-VR lap, participants filled out the Igroup Presence Questionnaire (IPQ) \cite{Schubert2003, Schubert2001, Regenbrecht2002}, the Simulator Sickness Questionnaire (SSQ) \cite{Kennedy1993}, and the NASA Task Load Index (NASA-TLX) \cite{Hart1988}, along with custom 5-point Likert ratings of perceived safety, comfort, and enjoyment. The without-VR lap provided a baseline for walking speed.

\subsection{Results}
As shown in Figure~\ref{fig:eval_res}, participants reported a high sense of spatial presence ($M=22.1/30$) and involvement ($M=17.7/24$), with lower realism scores ($M=12.6/24$), suggesting room for designing more detailed environments. Perceived safety ($M=3.9/5$) and comfort ($M=3.9/5$) were rated positively and were consistent across all lighting conditions, indicating robustness outdoors. Enjoyment ratings were moderate ($M=3/5$), reflecting opportunities for improving virtual content.

Walking speed was slower in the with-VR condition ($M=0.78$ m/s) than baseline ($M=1.08$ m/s). Participants required an average of $4.6$ (SD=1.15) course corrections over the route (500 m) in the with-VR condition. All participants completed the walk safely. SSQ scores were low (total $M=5.4$) and no participant reported any motion sickness.

\subsection{Discussion}
The high perceived safety ratings, irrespective of lighting conditions, suggest that \ww~\cite{nargund2024wavewalker} is suitable for outdoor VR use. The low SSQ scores and the absence of reported motion sickness indicate that the system effectively supports walking in VR while maintaining user comfort. The decrease in walking speed relative to baseline is likely due to participants being cautious while wearing a VR headset, consistent with prior findings \cite{mohler2007gait}, as well as increased dwell time associated with visual exploration of the virtual environment. 

\section{User Study II: Visualization Experiment}
\label{section:ve}

With Study~I indicating that the system is sufficiently safe and comfortable to use, we proceeded to evaluate how obstacle representation affects the user experience. Study~II compares three visualization techniques for static and dynamic obstacles to understand their impact on presence and walking behavior. This study provides the first comparative evaluation of obstacle visualizations in a camera-free outdoor VR system.  

\subsection{Participants}
A priori power analysis (medium effect size $f = 0.25$, $\alpha = 0.05$, $1 - \beta = 0.8$, correlation among repeated measures = $0.7$) indicated a sample of 18 would be sufficient. We recruited a different set of 18 participants (9 male, 9 female; ages 18–23, median = 20) who reported moderate familiarity with VR ($M = 3.2$ on a 5-point Likert scale). None had participated in Study~I. To enhance ecological validity, we did not explicitly control lighting conditions, resulting in ten participants completing the study in non-ideal lighting (very bright or dark) and eight in favorable lighting conditions.  
All participants provided informed consent, received \$15 compensation, and the study was approved by our local IRB (protocol \#anonymous).  

\subsection{Procedure}
Each participant walked a 200 m segment of the campus quad (yellow dashed line in Figure~\ref{fig:ve_route}) three times, once under each visualization condition, with order counterbalanced using a Latin-square design. A human chaperone accompanied all walks for safety, as in Study~I. The 200 m distance was selected to ensure participants had sufficient exposure (approximately 3-4 minutes per trial) to each visualization and to increase the likelihood of naturally occurring encounters with bystanders. 

Before the experiment, participants underwent a narrative induction where they were briefed on their objective to traverse a hostile alien landscape in order to reach their spaceship. They then completed a two-minute tutorial in VR. The tutorial demonstrated how obstacles would appear by using the researcher chaperone as an obstacle to ensure participants could recognize them during the study. After experiencing each condition, they completed the Igroup Presence Questionnaire (IPQ) \cite{Schubert2003, Schubert2001, Regenbrecht2002}, the NASA-TLX \cite{Hart1988}, the Cross-Reality Interaction Questionnaire (CRIQ) \cite{gottsacker_diegetic_2021}, and a custom survey adapted from the Collision Anxiety Questionnaire (CAQ) \cite{ring_caq_2024} as shown in Tables \ref{tab:ps_questions} and \ref{tab:ca_questions}. These instruments together captured spatial presence, task load, awareness, and collision-related anxiety. At the end of the study, participants indicated their preferred visualization condition and provided open-ended feedback on perceived safety and user experience.  

\begin{table}[!t]
    \begin{tabularx}{0.99\linewidth}{l X}
        \toprule
         1. & Did you notice the dynamic obstacles in a timely manner? \\
         2. & I was afraid of colliding a person who was moving \\
         3. & I was afraid of colliding with a person who was standing/sitting (static) \\
         4. & I felt uncomfortable because I could not judge where the other people around me were \\
         5. & I felt uncomfortable because I could not judge where I was in the real world \\
         6. & I knew at all times how far I was away from the other people around me \\
         7. & How safe did you feel like walking? \\
         \bottomrule
    \end{tabularx}
    \caption{The custom Perceived Safety questionnaire used in Study~II. Responses were captured on a 5-point Likert-type scale (1=\textit{Never}, 5=\textit{Always}). The items were adapted from the Collision Anxiety Questionnaire (CAQ)~\cite{ring_caq_2024} to assess participant's awareness and feeling of safety during outdoor VR locomotion.}
    \label{tab:ps_questions}
\end{table}

\begin{table}[h!]
    \begin{tabularx}{0.99\linewidth}{l X}
        \toprule
         1. & I was afraid of colliding with a person who was moving \\
         2. & I was afraid of colliding with a person who was standing/sitting (static) \\
         3. & I felt uncomfortable because I could not judge where the other people around me were \\
         4. & I felt uncomfortable because I could not judge where I was in the real world \\
         5. & I focused on the locations of the other people instead of the application \\
         6. & I knew at all times how far I was away from the other people around me \\
         7. & I held back or scaled down my movements while walking to avoid collisions\\
         \bottomrule
    \end{tabularx}
    \caption{The custom Collision Anxiety questionnaire used in Study~II. Responses were captured on a 5-point Likert-type scale (1=\textit{Never}, 5=\textit{Always}). The items were adapted from the Collision Anxiety Questionnaire (CAQ)~\cite{ring_caq_2024} to assess participants' collision related anxiety during outdoor VR locomotion.}
    \label{tab:ca_questions}
\end{table}

\subsection{Results}
We report results for perceived safety, collision anxiety, spatial presence, and task load, followed by participant preferences and qualitative feedback.

\subsubsection{Data Analysis}
A Shapiro-Wilk test was conducted on all dependent variables to assess normality. We used a repeated measures ANOVA when the assumption of normality was met and a Friedman test otherwise. All post-hoc pairwise comparisons were performed using Bonferroni corrections.   

\subsubsection{Obstacle and Localization Metrics}
Across all conditions, each participant on average encountered 26.3 ($\sigma=5.28$) dynamic obstacles (e.g., pedestrians, bikers) within the radar's field of view. The system successfully detected and rendered visualizations for over 96\% of these instances. The overall miss rate was 3.3\%, with most misses corresponding to objects moving at speeds greater than 8 miles per hours at the edge of the radar's range ($8 m$) or FoV ($\pm 70^{\circ})$. User localization was also stable with an average cumulative error of $1.64m$ ($\sigma = 0.93$) over the 200 m walking path and an average update interval of $0.051s$ ($\sigma = 0.0015$). 


\subsubsection{Perceived Safety}
Participants generally reported high perceived safety across all visualization conditions (Figure~\ref{fig:safety}). A repeated-measures ANOVA showed a significant difference for the question: \textit{``Did you notice dynamic obstacles in a timely manner?''} ($F = 4.63$, $p = 0.017$). Post-hoc tests indicated that point clouds enabled significantly faster detection of dynamic obstacles compared to alien avatars ($p = 0.043$), with no significant differences between point clouds and human avatars, or between the two avatar conditions (diegetic and non-diegetic). 

\subsubsection{Collision Anxiety}
Collision anxiety scores were low across all conditions (Figure~\ref{fig:collision}), and no significant differences were observed. 

\subsubsection{Spatial Presence}
\begin{figure*}[!t]
\centering
\begin{subfigure}{0.32\textwidth}
\includegraphics[width=0.9\linewidth]{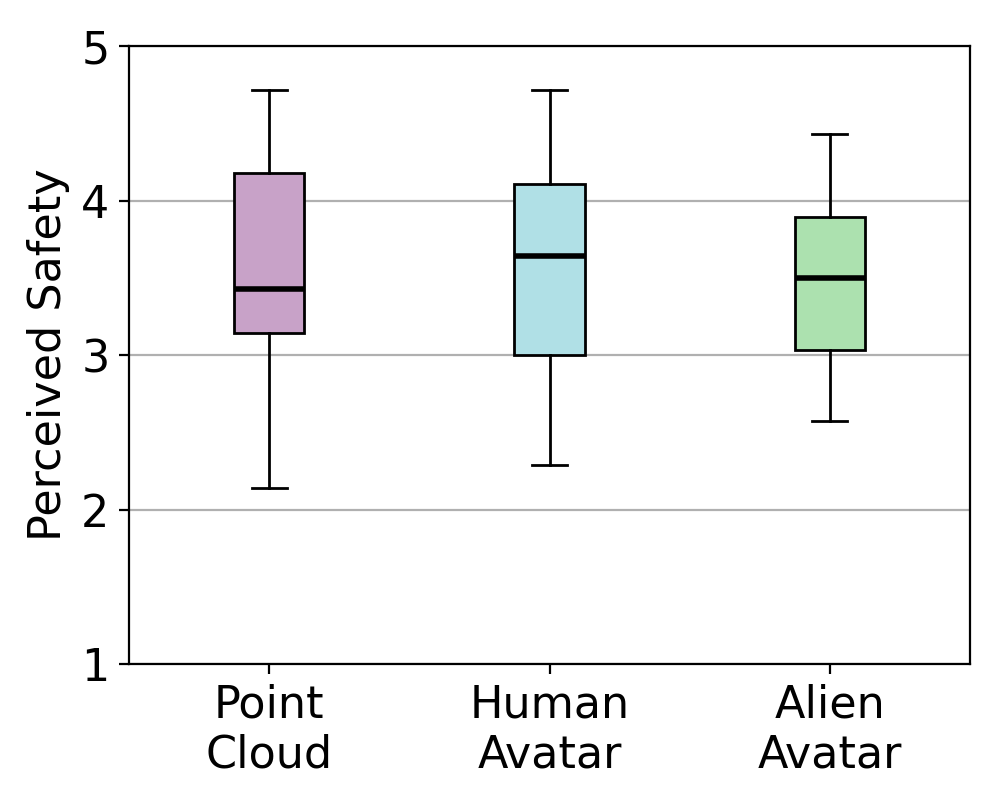} 
\caption{Perceived safety score}
\label{fig:safety}
\end{subfigure}
\begin{subfigure}{0.32\textwidth}
\includegraphics[width=0.9\linewidth]{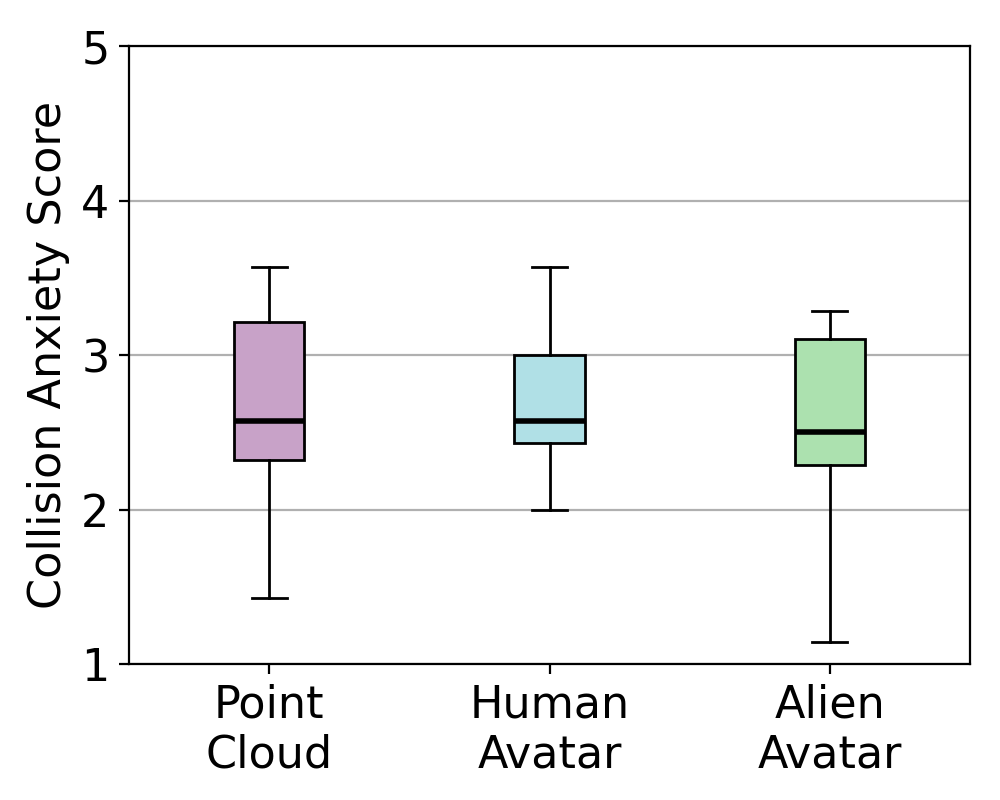}
\caption{Collision anxiety score}
\label{fig:collision}
\end{subfigure}
\begin{subfigure}{0.32\textwidth}
      \centering
\includegraphics[width=0.9\linewidth]{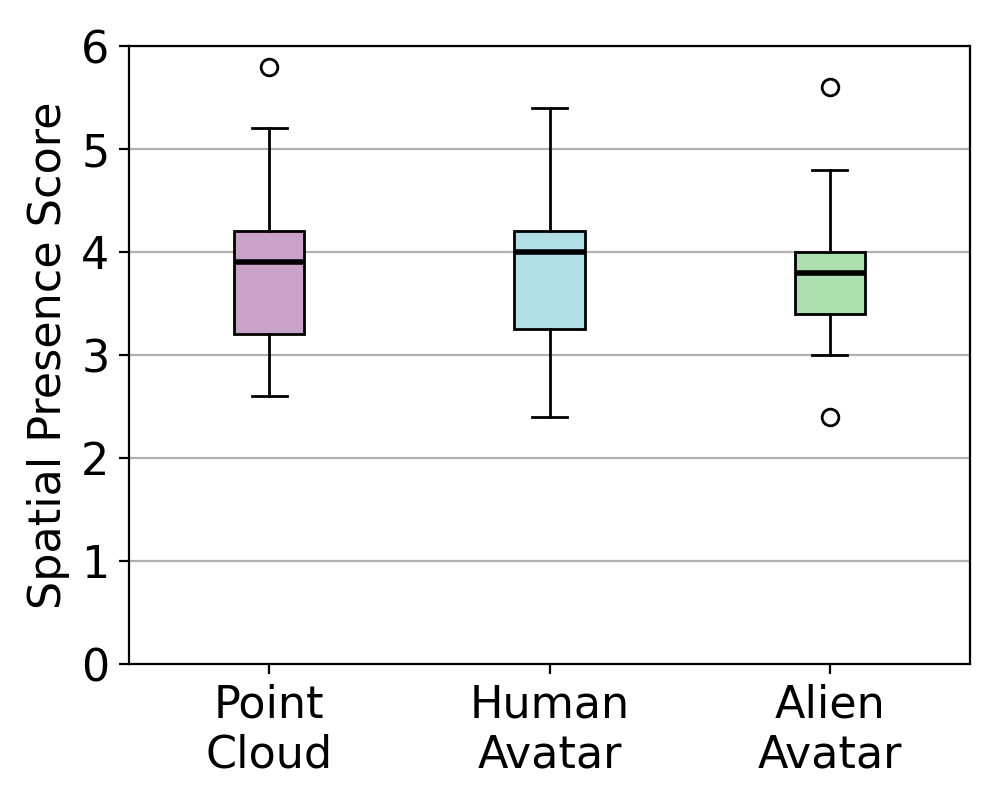}
    \caption{Perceived spatial presence level}
    \label{fig:presence}
\end{subfigure}
\caption{To understand user perceptions of the visualizations used in Study II, we measured three key experimental factors.  This figure shows participant scores for (a) Perceived Safety, (b) Collision Anxiety, and (c) Spatial Presence for each of our three visualization conditions: diegetic alien avatars, non-diegetic human avatars, and abstract point clouds.}
\label{fig:safety_collision}
\end{figure*}

Spatial presence scores were moderate and comparable across all conditions (Figure~\ref{fig:presence}). No statistically significant differences emerged, indicating that the choice of visualization did not lead to detectable variations in. perceived presence. 

\subsubsection{Task Load}
While the overall NASA-TLX scores showed no significant differences between visualization conditions ($F=0.59$, $p=0.54$, $\eta_{P}^{2}=0.03$), the scores for Effort and Frustration exhibited contrasting patterns with medium-to-large effect sizes. 
Effort was rated lowest for diegetic alien avatars ($M=30.3$), followed by non-diegetic human avatars ($M=37.5$) and abstract point clouds ($M=41.1$) ($F=2.86$, $p=0.071$, $\eta_{P}^{2}=0.14$). Frustration followed an inverse pattern: lowest for abstract point clouds ($M=17.8$), then non-diegetic human avatars ($M=19.2$), and highest for diegetic alien avatars ($M=26.9$) ($F=2.62$, $p=0.08$, $\eta_{P}^{2}=0.13$). While these sub-scale differences did not meet the threshold for statistical significance $(p < 0.05)$, the medium-to-large effect sizes suggest that a larger sample size might reveal a trade-off between immersion and cognitive simplicity.

\subsubsection{Preferences and Feedback}
Despite similar quantitative scores, participants expressed distinct preferences. Twelve of eighteen preferred the diegetic alien avatar, citing its narrative fit and smoother motion. However, some found it ambiguous, unsure whether it was interactive. Abstract point clouds were described as ``clear but abstract,'' valued for safety but seen as less immersive. Non-diegetic human avatars were rarely preferred, with several participants noting they felt ``out of place'' in the alien-themed environment. These perceptions echo prior findings that users favor diegetic cues for spatial awareness tasks \cite{gottsacker_diegetic_2021}.

Overall, the results indicate that while all visualization styles supported safe navigation, subtle differences in effort, frustration, and user preference suggest meaningful trade-offs between clarity, immersion, and cognitive load.

\subsubsection{Qualitative Results}

Participants provided written feedback after each condition, which we coded using inductive content analysis \cite{kyngas_inductive_2020}. Coding was performed by two researchers, who resolved disagreements through discussion. Responses clustered into three main safety-related themes and one set of visualization preferences.

\paragraph{Safety}
Although participants generally felt safe, occasional lapses were attributed to three issues:

\textbf{Missed Obstacles:} Several participants (P2, P3, P10, P12, P16, P18, P20) reported moments of discomfort when they could hear nearby bystanders but did not see a corresponding representation in VR. This sensory mismatch was a source of discomfort. For example, one participant noted, \textit{``...I wasn't able to see them but knew they were close around, that may have caused a bit of discomfort.''} (P10, Diegetic Alien Avatar). Another explained they felt unsafe, \textit{``Only when hearing people around me and no(t) seeing aliens move accordingly''} (P2, Diegetic Alien Avatar). 

\begin{figure}[!t]
    \centering
    \includegraphics[width=\columnwidth]{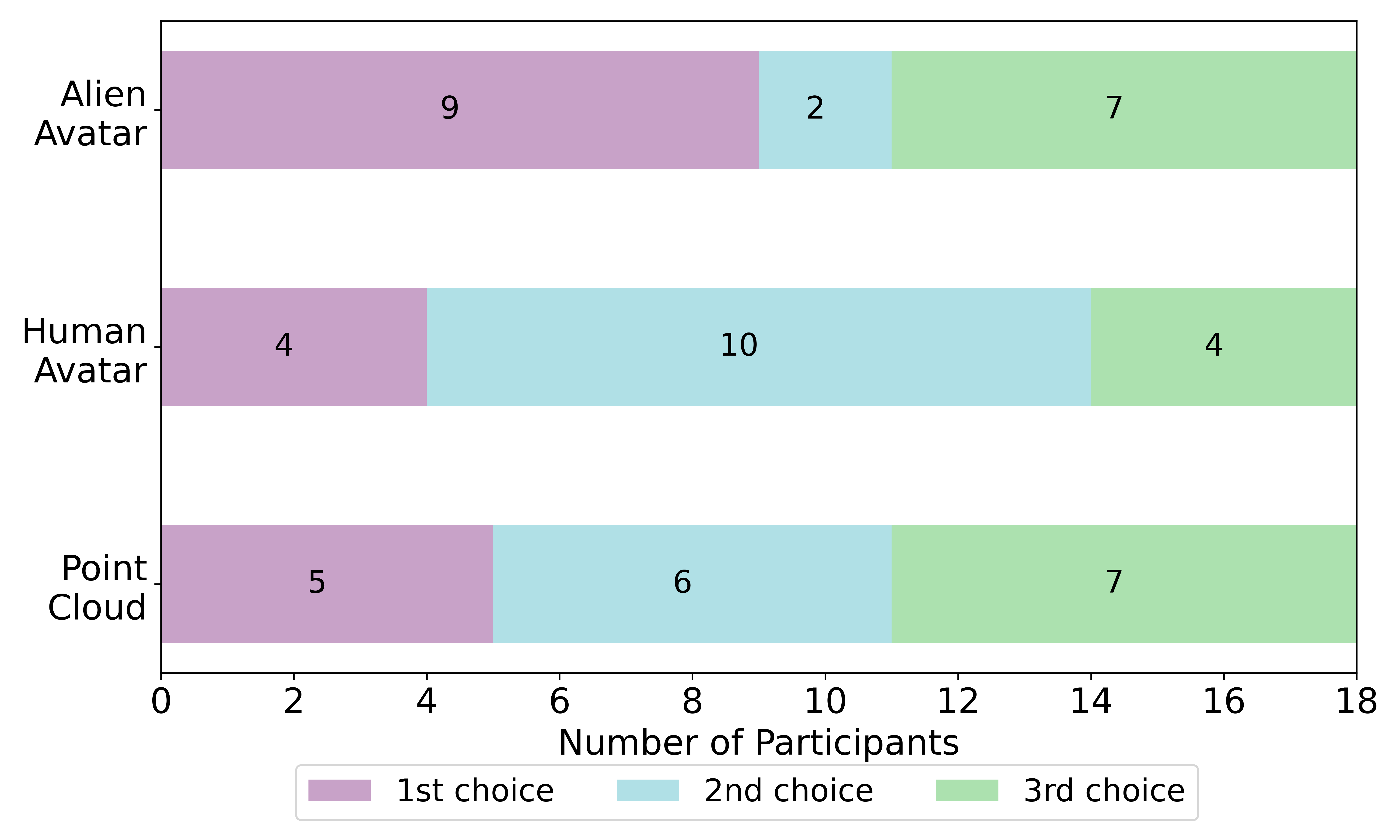}
    \caption{Participant preferences for the three visualization types are illustrated. The Diegetic Alien Avatar was selected as the first choice by 9 participants, while the Non-Diegetic Human Avatar was ranked second by 10 participants. The Abstract Point Cloud showed a more balanced distribution across rankings, with 7 participants identifying it as their least preferred option.}
    \label{fig:user_pref}
\end{figure}

\textbf{Ghost Tracks:} Multipath radar reflections occasionally generated ``ghost'' obstacles that ``popped'' into the scene. Ghost tracks were reported by 7 participants (P5, P6, P13, P14, P15, P18, P20), occurring in 8 of the 54 total trials. One participant described, \textit{``...the alien was right in front of me and I had to stop, but in reality the person was not that close''} (P20, Diegetic Alien Avatar). Another participant, while experiencing a ghost track, expressed that ``someone ran right in front of [them]'' (P5, Non-Diegetic Human Avatar). 

\textbf{Location and Extent of Obstacles:} Participants found avatars helpful for judging obstacle location but noted that point clouds better conveyed the size or density of groups. For example, one participant observed that avatars made it easier to gauge proximity (\textit{``a better grasp of how close they were''}, P19, Non-Diegetic Human Avatar), while another stated that point clouds made it easier to see when many people were clustered (\textit{``I could tell if there were a bigger congregation of people''}, P11, Abstract Point Cloud). 

\paragraph{Visualization Preferences}
When asked to rank conditions, participants most often selected the diegetic alien avatar visualization as their top choice (9 of 18), followed by abstract point cloud (5 of 18) and non-diegetic human avatar (4 of 18). Diegetic Alien Avatars were valued for narrative consistency, though they were rarely ranked second (2 of 18), indicating a polarized response. Abstract Point clouds were preferred for clarity but criticized for lack of immersion, and non-diegetic human avatars were frequently described as incongruent with the alien setting. While performance and safety scores were similar across conditions, subjective preference leaned toward diegetic visualization. 

These results indicate that while all three visualization conditions maintained safety, they offer distinct trade-offs between functional clarity and narrative immersion. 

\section{Discussion and Design Considerations}
Our findings present some implications for the design of obstacle visualizations in camera-free outdoor VR systems. Here we discuss how visualization choices need to balance the need for a user's sense of safety with the goal of creating an immersive, comfortable, and engaging user experience and present related design considerations.

\subsection{Perceived Safety and Comfort}

In outdoor VR, perceived safety is shaped by both tracking reliability and the way obstacles are visualized. While participants consistently reported feeling safe, our findings suggest that safety was not the product of any single factor but emerged from the interaction between tracking stability, sensor coverage, and representational clarity. 

\textbf{Path Deviation.}
Even minor drifts between virtual and physical paths undermined confidence in the system's spatial grounding. These lapses illustrate how safety in outdoor VR depends less on absolute positional accuracy and more on whether the system affords timely correction before deviations grow disruptive.

\textbf{Sensor Coverage.}
Although a single radar supported basic safety, its limited field of view and susceptibility to multipath signal reflections introduced moments of doubt. This points to a trade-off that minimal hardware can sustain functionality, but broader coverage and redundancy are likely needed to reinforce trust in dynamic, real-world environments.

\textbf{Representational Roles.}
Different visualization styles offered complementary safety cues. Avatars, for instance, were effective at conveying the position and trajectory of individual obstacles. In contrast, point clouds better signaled density and extent of obstacles and groups. This trade-off helps explain a key quantitative finding: while overall safety ratings were similar, point clouds scored significantly better for the timely detection of dynamic obstacles. This was likely because the raw spatial data of point clouds was more effective at preventing missed detections or sensory conflicts, even if avatars were better for interpreting the movement of a single, detected object. The lack of significant differences in overall ratings suggests that users ultimately leveraged whichever cues were most salient in context, pointing to the value of hybrid or flexible designs.

\textbf{Collision Anxiety.}
The absence of elevated collision anxiety across conditions highlights the importance of reliability and consistency over stylistic differences. As long as users could depend on consistent obstacle signaling, they maintained confidence in navigating naturally without hesitation.

\textbf{Design Implications.}
These patterns suggest several directions for enhancing safety in camera-free outdoor VR. Subtle guidance features (e.g., ground ribbons or peripheral drift indicators) can help users self-correct alignment errors. Opportunistic radar-anchored corrections and multi-radar fusion may address ghost artifacts while preserving privacy. Hybrid visualizations that pair avatars for localization with abstract overlays for extent could combine immersion with clarity. In high-density or uncertain environments, layering urgency cues such as predictive trails or intensity gradients may help sustain trust.

\subsection{Sense of Presence}

Our findings suggest that in outdoor VR, a user's sense of presence may depend less on the visual style of obstacle representations and more on the stability and coherence of the sensing–display loop. Contrary to our expectations, diegetic avatars did not yield stronger presence than other visualization conditions, pointing instead to the primacy of technical fidelity and cross-modal consistency.

\textbf{Tracking Delay.}
Even modest sensing delays undermined the impression of a stable environment. These moments highlight that presence is not only a function of visual design but also of how reliably the system maintains synchrony between bodily movement and environmental response.

\textbf{Avatar Movement.}
When orientation estimates, inferred from obstacle movement, were noisy avatars would occasionally move in implausible ways. Such breaks in behavioral realism weakened plausibility, illustrating how presence relies on consistent mappings between user expectations and system output rather than on visual richness alone. Prior work links such inconsistencies to reduced plausibility and engagement \cite{toczek_influence, jung_perspective_2021}.

\textbf{Sensory Conflicts.}
Cross-modal mismatches, such as hearing bystanders without corresponding visual cues or seeing ghost tracks with no audible presence, disrupted the coherence of the environment. These inconsistencies reveal how presence is shaped by the alignment of sensory channels as much as by the fidelity of any single modality. These cross-modal inconsistencies are known to disrupt presence by violating embodied expectations and multisensory integration \cite{grassini_evaluating_2021, pfeiffer_multisensory_2013}. 

\textbf{Visualization Style.}
The lack of significant differences in presence across visualization types suggests that obstacle representations functioned more as perceptual cues than as immersive narrative elements. In this context, consistent and reliable signaling may have outweighed any benefits of diegetic or stylistic coherence, echoing patterns seen in collision anxiety responses.    

\textbf{Design Implications.}
Supporting presence in outdoor VR requires prioritizing the coherence of the sensing–display pipeline in addition to visualization techniques. Reducing latency through opportunistic calibration and introducing radar-based localization updates can help maintain spatial consistency. When orientation is uncertain, abstract representations may be preferable to avatars that imply precision, while smoothing or predictive filters can preserve plausibility. Subtle multimodal cues can help reconcile discrepancies across senses. Visualization style should therefore be understood as augmenting and not substituting for technical stability and multisensory coherence in presence-critical VR contexts.

\subsection{Semantic-Behavioral Alignment}
A critical design challenge emerged regarding the alignment between the semantic cues embedded in the visualization and the behaviors users inferred from them. For example, the Alien avatars were designed to be consistent with the internal narrative logic of the VR environment. However, their visual form and styling may have communicated expectations about interactability that were not implemented.

By visualizing pedestrians, cyclists, and other dynamic obstacles as weapon-carrying aliens near the user may have signaled potentially interactive behavior (e.g., aggression, intent). However, the avatars behaved passively, navigating around the user without exhibiting intent or interaction. This mismatch between signaled meaning and observed behavior constitutes an expectation violation that likely produced cognitive dissonance~\cite{hameedTaxonomy2024} and may have contributed to a break in presence~\cite{skarbez2016}.

By contrast, the Point Cloud condition offered high functional clarity and minimized semantic signaling. While this reduced the risk of expectation mismatch, participant feedback indicates that it introduced interpretive overhead. The lack of recognizable form required users to actively infer obstacle identity and spatial properties, such as distance, through additional cognitive mapping. Taken together, these findings highlight a trade-off between semantic richness and behavioral legibility.

\textbf{Design Implications.}
This suggests a fundamental trade-off in the design of camera-free outdoor VR systems. Increasing the thematic congruence of obstacle representations may obscure their real-world identity as bystanders, thereby confusing users when visual semantics suggest interactability. Therefore, when using diegetic representations, designers should consider avoiding cues that imply interactability, as these cues may prompt users to approach obstacles instead of maintaining a safe distance. This could be achieved through restrained visual styling, consistent overlays, or subtle cues that emphasize passive functional status. Hybrid approaches that layer minimal abstract markers (e.g., halos or ground anchors) onto diegetic avatars may preserve immersion while explicitly communicating that the avatar corresponds to a real-world bystander. Ultimately, visualization strategies should not be evaluated only for their visual fidelity, but for how effectively they manage user expectations about the obstacle.

\subsection{Representation Design}

Preferences for obstacle representations were shaped by both functional clarity and narrative integration. The diegetic alien avatar, while thematically consistent with the VR environment, elicited polarized responses where it was ranked highest by half of participants, but also tied for the most last-place rankings. This suggests that high diegetic integration can enhance immersion for some users while creating confusion for others regarding an object's purpose. The non-diegetic human avatar, although visually familiar, often felt incongruent within the alien world and was rarely preferred. In contrast, the point cloud was valued for its functional clarity and unambiguous role but offered limited immersion, leading to moderate and divided rankings. Participant preference for the diegetic visualization aligns with prior findings that users favor diegetic cues for awareness tasks \cite{gottsacker_diegetic_2021}, despite modest quantitative differences across conditions.

These results highlight the challenge of designing representations that are both intuitively interpretable and narratively coherent. Overly diegetic forms may be misinterpreted as interactive or agentive, while abstract visualizations, though clearer in function, can reduce engagement. Participants expressed interest in more flexible or customizable visualization options such as the ability to toggle between symbolic indicators (e.g., outlines or icons) and immersive avatars, depending on context and personal preference. Hybrid designs that blend recognizable human or symbolic cues with environment-matching aesthetics may better balance clarity and immersion.

Taken together, our findings suggest that visualization design in outdoor VR must negotiate two interdependent goals of {\bf functional clarity} for safe navigation and {\bf narrative integration} for immersive engagement. Across both studies, all visualization types supported safe use, but subtle differences in presence, effort, frustration, and preference revealed trade-offs between diegetic and abstract approaches. No single strategy proved universally optimal, indicating that adaptable or hybrid designs may best accommodate diverse user needs and dynamic outdoor environments. For designers, this highlights the importance of coupling reliable sensing with flexible, context-aware representations that prioritize safety while enhancing comfort and immersion.

\section{Limitations and Future Work}
While our studies provide insights into the design of obstacle visualizations for outdoor VR, several limitations suggest directions for future work. 

We did not measure objective safety metrics such as false alarm and miss rates, end-to-end latency, or the influence of head movement on object detection. Future studies could systematically characterize obstacle detection and tracking performance, for example using ROC analyses or latency benchmarks, to complement the subjective safety reports collected here. Both studies took place on a campus quad, a setting representative of public pedestrian spaces but lacking features common in parks or urban streets, such as pets, uneven terrain, or vehicular traffic. These factors may affect sensing reliability and user behavior. Future work should evaluate outdoor VR systems in more varied environments to understand generalizability. Occasional GPS noise and radar reflections introduced delays, missed detections, and ghost tracks, particularly near tall buildings. While higher-precision GPS/IMU modules, additional radars, or sensor fusion approaches could mitigate these issues, the use of cameras raises privacy concerns that warrant careful consideration. Exploring privacy-preserving multimodal sensing remains an important direction.

To avoid masking real-world cues, our implementation excluded soundscapes. However, participants occasionally reported uncertainty about obstacles outside the radar's field of view. Adding rear-facing radars paired with subtle auditory cues or ambient soundscapes may improve spatial awareness and immersion \cite{cummings_how_2016, bosman_effect_2024}. Future work should examine whether such cues enhance presence without introducing distraction or misinterpretation.

While our visualizations were within the field of view, the virtual environment had few interactive or narrative elements. These factors may have reduced perceptual clarity and limited differentiation between visualization conditions. Increasing the salience of visualizations, enriching the virtual scene with additional agents or interactivity, or enhancing graphical fidelity could help amplify differences in presence and user preference.

Together, these limitations suggest that while our findings indicate the feasibility of camera-free outdoor VR and the promise of hybrid visualization strategies, further work is needed to validate safety at scale, extend evaluation to more complex environments, and refine both sensing and representational design.

\section{Conclusion}

Our studies suggest that camera-free outdoor VR using mmWave radar can support safe and immersive experiences over a wide range of lighting conditions, while preserving bystander privacy. In a preliminary system evaluation, participants reported feeling safe and comfortable across varied outdoor conditions. Building on this foundation, we compared three obstacle visualization strategies and found that while all conveyed spatial awareness effectively, each carried distinct strengths: avatars supported location and direction, point clouds emphasized extent, and diegetic integration shaped preference but also introduced ambiguity. From these findings, we distilled design considerations for balancing clarity and immersion in outdoor VR. Future systems may benefit from hybrid or customizable visualization approaches that adapt to user needs and environmental complexity. More broadly, this work indicates that reliable sensing coupled with thoughtful visualization design is central to extending VR beyond controlled indoor settings, toward everyday outdoor experiences that prioritize both safety and privacy.

\bibliographystyle{abbrv}

\bibliography{mmwave}
\end{document}